\documentclass{article}
\usepackage{PRIMEarxiv}
\usepackage{amsmath,graphicx,booktabs,tabulary,hyperref, subcaption}
\usepackage[linesnumbered,ruled,vlined]{algorithm2e}

\title{Cluster-to-Predict Affect Contours from Speech
}

\author{
  Gökhan Kuşçu, Engin Erzin \\
  KUIS-AI Laboratory \\
  Multimedia, Vision and Graphics Group \\
  College of Engineering, Ko\c{c} University, Istanbul, Turkey\\
  \texttt{\{gkuscu21, eerzin\}@ku.edu.tr} \\
}

\begin{document}
\maketitle
\begin{abstract}
Continuous emotion recognition (CER) aims to track the dynamic changes in a person's emotional state over time. This paper proposes a novel approach to translating CER into a prediction problem of dynamic affect-contour clusters from speech, where the affect-contour is defined as the contour of annotated affect attributes in a temporal window. Our approach defines a cluster-to-predict (C2P) framework that learns affect-contour clusters, which are predicted from speech with higher precision. To achieve this, C2P runs an unsupervised iterative optimization process to learn affect-contour clusters by minimizing both clustering loss and speech-driven affect-contour prediction loss. Our objective findings demonstrate the value of speech-driven clustering for both arousal and valence attributes. Experiments conducted on the RECOLA dataset yielded promising classification results, with F1 scores of 0.84 for arousal and 0.75 for valence in our four-class speech-driven affect-contour prediction model.
\end{abstract}
\section{Introduction}
\label{sec:intro}

As human-computer interaction becomes more integrated into everyday life, the research community places a strong emphasis on speech-emotion recognition (SER) research. SER shows promise in enhancing the natural exchange of communication between individuals and computer systems. The applications of SER extend across diverse domains, such as driver assistance, healthcare, entertainment, chatbots, and more. The knowledge derived from SER technology facilitates more authentic interactions and provides valuable insights into understanding human behavior for intelligent systems \cite{lee2023}. 

Categorical and dimensional affect models are commonly utilized \cite{scherer2008}. According to Ekman's categorical model \cite{ekman1992}, there are six primary emotions: anger, contempt, fear, happiness, sorrow, and surprise. The categorical model, which describes the classification of basic emotion categories, has been extensively studied in SER literature. In contrast, Russell's dimensional circumplex model expresses basic emotions on continuous dimensions of arousal (low vs high) and valence (unpleasant vs pleasant) attributes \cite{russell1980}. The 2D arousal-valence model has advantages over the categorical model, as it can describe continuously fluctuating intensity levels across categorical emotions.

SER models typically define the regression task of predicting emotional attributes (arousal, valence) and the classification task of predicting emotional categories (happy, sad, ...). While unweighted/weighted accuracy, F1 score, or similar metrics are used in assessing the classification task, the output of the regression task is assessed with correlation-based metrics to match the trends on the continuous ground truth and the regressed emotional dimensions. A recent survey of SER tasks on extensions of deep representation learning highlights the popularity of LSTM/GRU-RNNs combined with CNNs for supervised tasks, and Denoising Autoencoders (DAEs), Variational Autoencoders (VAEs), and GAN-based models for unsupervised representation learnings \cite{siddique2023}.

Dynamic temporal modeling appears to be another important aspect of the SER problem. A recent study presents a chunking-based approach for sentence-level information extraction from varied-length acoustic feature sequences to better cope with dynamic temporal modeling for the prediction of emotional dimensions \cite{lin2023}. 
Another study presents two architectures for capturing long-term temporal dependencies in acoustic features \cite{khorram2017}. Dilated convolutions maintain the input signal's length while incorporating long-term information via filters with varying dilation factors. Down/upsampling networks downsample the signal to grasp global features, then reconstruct the output to match the uncompressed input length. 
Deep neural networks are widely explored for SER tasks as well. In a recent study, Wu et.al explores neural architecture search (NAS) for automatically generating customized SER models \cite{meng2022}.

Some studies combine regression and classification problems of SER. In such a study, the regression problem of emotional attributes was translated into a classification problem by discretizing the training labels at different resolutions \cite{provost2017}. Then, a multi-task bidirectional LSTM network is trained to jointly predict label sequences at different resolutions, and an emotion decoding algorithm produces more robust continuous emotional attribute estimates.
In another study, AlBadawy and Kim propose a joint modeling method merging discrete and continuous emotion representations \cite{AlBadawy2018}. Ensemble and end-to-end approaches balance between discretized and continuous representations using D-BLSTM architectures. 

Some other studies are different from the typical regression and classification problems of SER as well. A recent study explores emotional similarity measurement by estimating the most similar emotional content among two alternatives to a given anchor \cite{harvill2023}. This task states a different SER problem that leads to learning representations of emotional similarity. Another variation of the SER task defines emotionally salient regions relying on the qualitative agreement of raters for arousal and valence and then uses an ensemble of BLSTM predictors to detect emotionally salient regions \cite{busso2018}.

Semi-supervised learning (SSL) has been used to resolve the generalization problem for SER through pseudo-labeling classification. DeepEmoCluster presents such an SSL-based framework to learn latent representations to define emotional clusters from pseudo-labeling classification \cite{busso2021}. In DeepEmoCluster, a CNN-based feature extraction forms latent representations from speech to perform emotional attribute regression, and a second network forms clusters of latent representations to assign pseudo-class labels via SSL. A recent study extends DeepEmoCluster by introducing sentence-level temporal modeling by integrating temporal constraints via temporal network and triplet loss function \cite{busso2024}.

%\subsection{Contribution of paper}

In this paper, our main motivation is to define a new SER problem, different from regression and classification tasks, which aims to discover temporal affect-contour clusters that are highly predictable from speech representations. Here, we define the affect contour as the contour of annotated affect attributes in a temporal window. To solve this problem, we propose a novel SSL approach, which defines a cluster-to-predict (C2P) framework. The proposed C2P framework is expected to discover novel affect contour clusters that are predictable from speech with higher precision. The main building blocks of the proposed C2P framework include i) the affect network, which extracts latent representations of affect contours and maps them to the affect clusters; ii) the speech network, which extracts latent representations of the speech signal and maps them to the affect clusters, and iii) the clustering block, which runs k-means clustering update on the latent representations of affect contours. Note that although the DeepEmoCluster and the proposed C2P frameworks have similar SSL approaches, the proposed C2P framework differs with key distinctions. The C2P framework defines the pseudo-labeled clusters from the latent representations of the affect contour but not from speech. Furthermore, the C2P framework learns to predict the affect-contour cluster labels from speech by defining a classification problem. Experimental results show the effectiveness of our approach in classification and provide insightful information on arousal-valence space through the discovered affect contour clusters.

\section{Methodology}
\label{sec:methodology}

In this study, we aim to discover temporal affect-contour clusters that are highly predictable from speech representations. We propose a novel SSL approach by defining a cluster-to-predict (C2P) network. 
%The proposed C2P network discovers novel affect-contour clusters that are predictable from speech with higher precision. 

The proposed C2P network consists of two main blocks: an affect network and a speech network. The speech network (SpeechNet) receives the speech signal as input and predicts affect-contour cluster labels. The affect network (AffectNet) receives emotional attribute vectors as input and employs two tasks. Firstly, AffectNet extracts a reduced dimensional latent affect representation and updates the \(k\)-means centroids defining the affect-contour cluster labels. Secondly, AffectNet predicts affect-contour cluster labels from the latent affect representations.

\subsection{SpeechNet}

The SpeechNet predicts affect-contour cluster labels from the speech input in two phases. The first phase includes a pre-trained acoustic feature extractor. The second phase maps high-dimensional acoustic feature embeddings into low-dimensional acoustic latent representations and then to the affect-contour cluster labels.

Wav2Vec is a powerful learned acoustic feature extractor for speech data \cite{wav2vec}. Unlike traditional hand-crafted acoustic features, Wav2Vec operates directly on the raw audio waveform to provide a compressed representation of the speech signal. We choose Wav2Vec as the pre-trained acoustic feature extractor, where the acoustic feature embeddings are taken from block 15 of the pre-trained Wav2Vec 2.0 \textit{Large} model as 1024-dimensional vectors for 20~ms frames. The choice of block 15 is motivated by its proven efficacy in decreasing phoneme error rates \cite{wav2vec}. The Wav2Vec block receives a temporal window of $N_s$ speech frames to construct $[N_s, 1024]$ dimensional acoustic feature embeddings for each temporal window. In this study, we chose temporal window duration as 2 seconds with $N_s=99$.

The second phase includes a multi-layer CNN network and a fully connected predictor (SpeechCNN). The CNN network extracts low-dimensional acoustic latent representations from the Wav2Vec-driven acoustic feature embeddings. The fully connected layer predicts the affect-contour cluster labels from the acoustic latent representations. The network architecture of the SpeechCNN predictor is given in Table~\ref{tbl:SpeechCNN}.

\begin{table}[bht]
\caption{Network architecture of the SpeechCNN predictor}
\label{tbl:SpeechCNN}
\centering
\scalebox{0.90}{
\begin{tabular}{l c c c c}
\toprule[1pt]\midrule[0.3pt]
\textbf{Layer} & \textbf{Type} & \textbf{Depth} & \textbf{Kernel} & \textbf{Output}\\
\midrule
1 & Wav2Vec & - & - & 99x1024x1 \\
\hline
2 & CONV+ReLU & 32 & 7x7 & 93x1018x32 \\
\hline
3 & MaxPooling & 32 & 2x2 & 46x509x32 \\
\hline
4 & CONV+ReLU & 64 & 7x7 & 40x503x64 \\
\hline
5 & MaxPooling & 64 & 2x2 & 20x251x128 \\
\hline
6 & CONV+ReLU & 128 & 5x5 & 16x247x128 \\
\hline
7 & CONV+ReLU & 256 & 5x5 & 12x243x256 \\
\hline
8 & MaxPooling & 256 & 2x2 & 6x121x512 \\
\hline
9 & CONV+ReLU & 512 & 3x3 & 4x119x512 \\
\hline
10 & CONV+ReLU & 512 & 3x3 & 2x59x512 \\
\hline
11 & MaxPooling & 512 & 2x2 & 1x29x512 \\
\hline
12 & FC+LeakyReLU & - & - & 1024 \\
\hline
13 & FC+LeakyReLU & - & - & 512 \\
\hline
14 & FC+LeakyReLU & - & - & 128 \\
\hline
15 & FC+LeakyReLU & - & - & 64 \\
\hline
16 & FC+Softmax & - & - & 4 \\
\bottomrule[1pt]\midrule[0.3pt]
\end{tabular}}
\end{table}

\begin{table}[h]
\caption{Network architecture of the AffectNet model}
\label{tbl:AffectNet}
\centering
\scalebox{0.90}{
\begin{tabular}{l c c c c}
\toprule[1pt]\midrule[0.3pt]
\textbf{Layer} & \textbf{Type} & \textbf{Depth} & \textbf{Kernel} & \textbf{Output}\\
\midrule
1 & CONV+ReLU & 32 & 3 & 32 \\
\hline
2 & CONV+ReLU & 16 & 3 & 16 \\
\hline
3 & CONV+ReLU & 8 & 3 & 8 \\
\hline
4 & FC & - & - & 4 \\
\bottomrule[1pt]\midrule[0.3pt]
\end{tabular}}
\end{table}

\begin{figure}[t]
  \centering
  \includegraphics[width=0.90\linewidth]{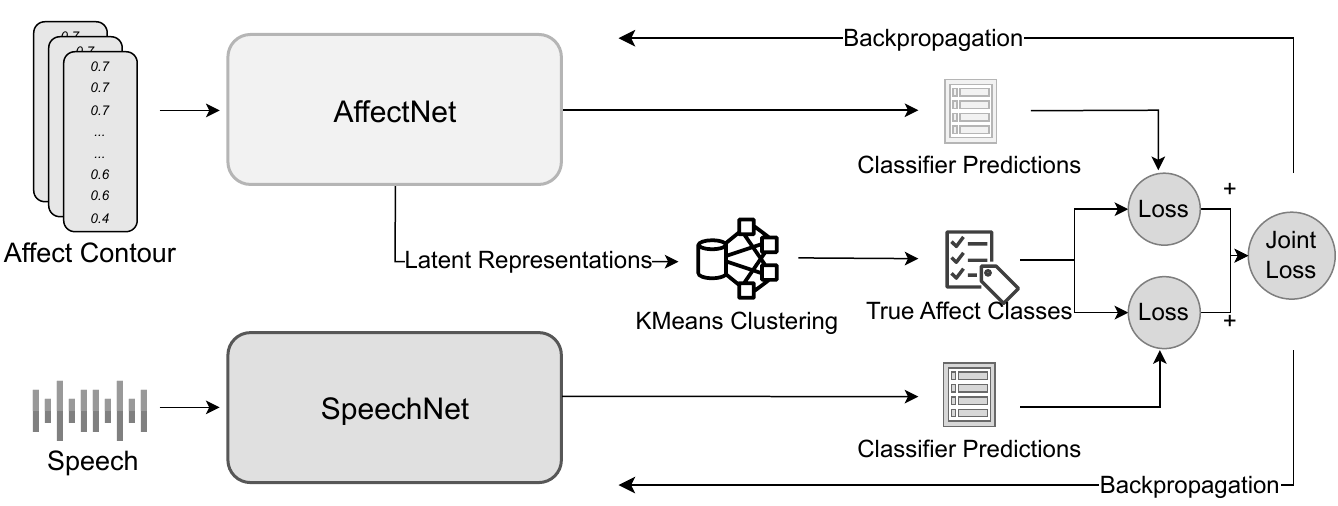}
  \caption{Block diagram of the proposed C2P network}
  \label{fig:speech_production}
\end{figure}

\subsection{AffectNet}

Similar to the construction of the speech network, our affect clustering network (AffectNet) is designed as a convolutional neural network with fully connected layers responsible for predicting affect classes. This network operates with dual objectives. The primary objective involves acting as an autoencoder \cite{transformingautoencoder} without the decoder component. The absence of the decoder is compensated by employing the classification part to compute the loss and update the weights. This design compels the convolutional network to generate features that enhance classification performance. The first target is achieved by extracting a lower-dimensional representation of the affect contour by mapping the $N_a$-unit long affect contour into informative latent affect representations. With the 2~second temporal window setting, we take $N_a=50$. These latent representations are subsequently clustered into distinct affect classes through \(k\)-means clustering. Simultaneously, as a secondary objective, the entire network functions as a classifier, predicting the affect-cluster labels. The loss function, comprising cross-entropy loss for multi-class classification, is aggregated with the speech network loss. Subsequently, the weights of both networks are updated accordingly. Given the task definition of the affect network, the weight convergence occurs at updated points that enhance the model's ability to produce superior latent representations of the affect contour, facilitating improved clustering in subsequent training iterations.

The AffectNet is constructed by concatenating convolutional networks with the incorporation of dropout layers as given in Table~\ref{tbl:AffectNet}. Notably, the kernel sizes of the convolutional layers in the earlier segments gradually diminish towards the network's last layers. This deliberate reduction is implemented to ensure a broader receptive field during the initial stages, facilitating the extraction of global features while simultaneously promoting the detailed extraction of local features.

Regarding the latent representations extracted for the clustering component, we investigated the impact of selecting intermediate layers and identified a trade-off. Deeper layers contain more informative features about the AffectNet; however, the lower-dimensional space proves less effective when applying the \(k\)-means algorithm. In seeking an optimal balance between these considerations, we select an 8-dimensional latent affect representation that performs well for both objectives. 
%Consequently, this dimension is chosen for experimental evaluations of the results.

\subsection{C2P Network}

The proposed C2P network is jointly defined by the AffectNet and SpeechNet networks. The training routine of the proposed C2P network is given in Algorithm~\ref{algo:c2p}.

The joint loss of the C2P network is defined as 
%\begin{equation}
    $L = \alpha L_a + (1-\alpha) L_s$,
%\end{equation}
where $\alpha$ is a weighting factor, and $L_a$ and $L_s$ are correspondingly cross-entropy losses of the AffectNet and SpeechNet. During the training, the network parameters and the \(k\)-means clusters of the latent affect representations are jointly updated in an alternating order. The latent affect representations are clustered once for echo epoch using the \(k\)-means clustering to maintain stability across batches. These clusters are applied to speech segments, which are then fed into the C2P network in batches. The network converges more smoothly by conducting the clustering step separately from the training loop and using the clustering centroids as initial points for subsequent epochs.

\begin{algorithm*}[hbt]
  \footnotesize
  %\scriptsize
  %\small % Adjust font size
  %\setstretch{0.8} % Adjust line spacing
  \caption{Training Routine of the Proposed C2P Network}
  \label{algo:c2p}
  \KwIn{{\it speech} signal, {\it affectCon} affect contour}
  \KwOut{Trained {\it AffectNet, SpeechNet} networks}
  %\tcp{Segment data, 2-second long speech segments with 1-second intercepts}
  {\it speSegs, affSegs} $\leftarrow$ SegSpeech({\it speech}), SegAffect({\it affectCon}); \tcp{Segment data}
  
  {\it AffectNet, SpeechNet} $\leftarrow$ InitAffectNet(), InitSpeechNet(); \tcp{Initialize models}
  %\tcp{Epoch Loop}
  \For{{\it epoch} $\leftarrow 1$ \KwTo {\it totalEpochs}}{ 
    {\it \_ , latentReps} $\leftarrow$ AffectNetForward({\it affSegs}); \tcp{Latent representations from frozen AffectNet}
    
    {\it clusters} $\leftarrow$ KMeansClustering({\it latentReps}); \tcp{Latent representations to affect clusters}

   % \tcp{Training Loop}
    \For{{\it batch} $\leftarrow 1$ \KwTo {\it totalBatches}}{
      
      {\it affectPreds, \_} $\leftarrow$ AffectNetForward({\it affSegs[batch]}); \tcp{AffectNet forward}
      
      {\it speechPreds} $\leftarrow$ SpeechNetForward({\it speSegs[batch]}); \tcp{SpeechNet forward}
      
      {\it totalLoss} $\leftarrow$ CalculateTotalLoss({\it affPreds, spePreds, clusters}); \tcp{Calculate loss}
      
      Backpropagate{\it (AffectNet, SpeechNet, totalLoss)}; \tcp{Backpropagate}
    }
  }
  
  \Return{{\it AffectNet, SpeechNet}}; \tcp{Trained models}
\end{algorithm*}

\section{Experimental Evaluations}
\label{sec:experimental}

\subsection{Database Overview}

The RECOLA dataset \cite{6553805} is widely acknowledged as a significant resource extensively utilized in various multimodal emotion recognition tasks, prominently featured in AVEC challenges spanning different years, including \cite{avec16} and \cite{avec18}. It comprises a collection of 9.5 hours of multimodal recordings, incorporating audio, visual, and physiological data. These recordings capture real-time dyadic interactions involving 46 French-speaking participants collaboratively engaging in a task. The audio subset of the dataset, crucial for our study, consists of 27 five-minute speech utterances distributed into training, development, and test sets, each accompanied by corresponding affect contours. The audio modality and its associated affect contours for arousal and valence were annotated by six gender-balanced annotators at a frame rate of 40 milliseconds. This study adopts the mean value obtained from these six annotations as the reference point for our evaluation.

\subsection{Experimental Setup}

The whole dataset is split into training, development, and test parts, where each part consists of nine five-minute speech data segments concatenated to produce a total of 45~minutes of speech data. Similarly, the corresponding arousal and valence attributes form 1-dimensional vectors of 45~minutes. The data processing runs over 2-second temporal windows, which are sliding with 1-second intervals over time. Since the test part of the RECOLA dataset is not public, we run all training and testing evaluations on the training and development partitions of the RECOLA dataset, respectively.  

The C2P network is trained for various numbers of clusters $k=2, 2, ...,10$ both for arousal and valence attributes. For both attributes, the within-cluster variance is observed to reach a close-minimum value by $k=4$. Hence, we fixed the $k=4$ for both arousal and valence attributes in the experiments.
Furthermore, the weight for the joint loss is selected as $\alpha = 0.2$ by running a grid search to minimize the joint loss. 

The training utilizes the Adam optimizer with a learning rate of 0.001. It operates in batches of 256 samples across 50 epochs, with early stopping implemented to cease training if consecutive epochs do not show significant improvement beyond a certain threshold in terms of cumulative loss. Hyperparameters are adjusted iteratively during experimentation. 

\subsection{C2P Learned Affect Contours}

\begin{figure}[t]
  \centering
  \begin{subfigure}{0.49\textwidth}
    \includegraphics[width=\textwidth]{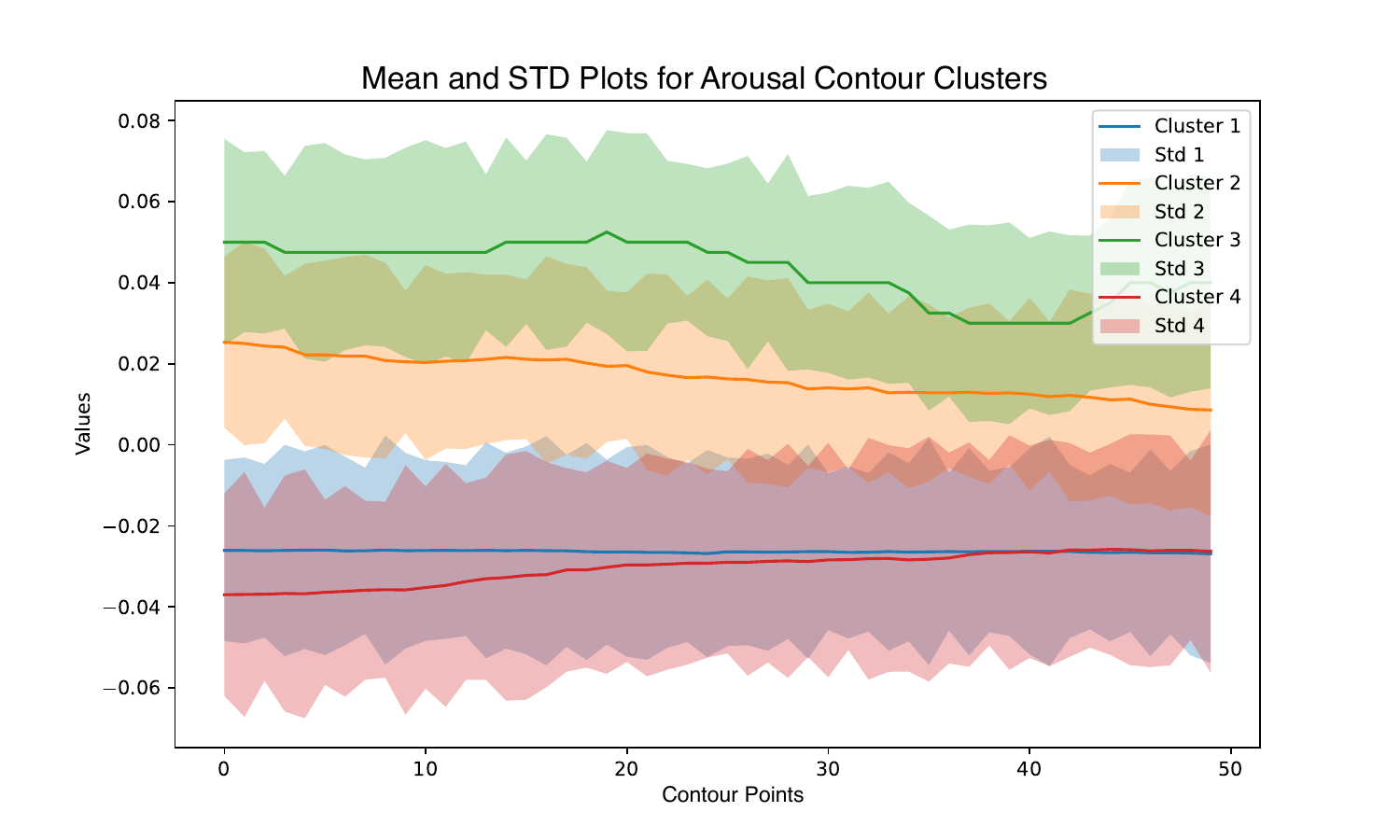}
    \caption{Arousal}
    \label{fig:first}
  \end{subfigure}
  \hfill
  \begin{subfigure}{0.49\textwidth}
    \includegraphics[width=\textwidth]{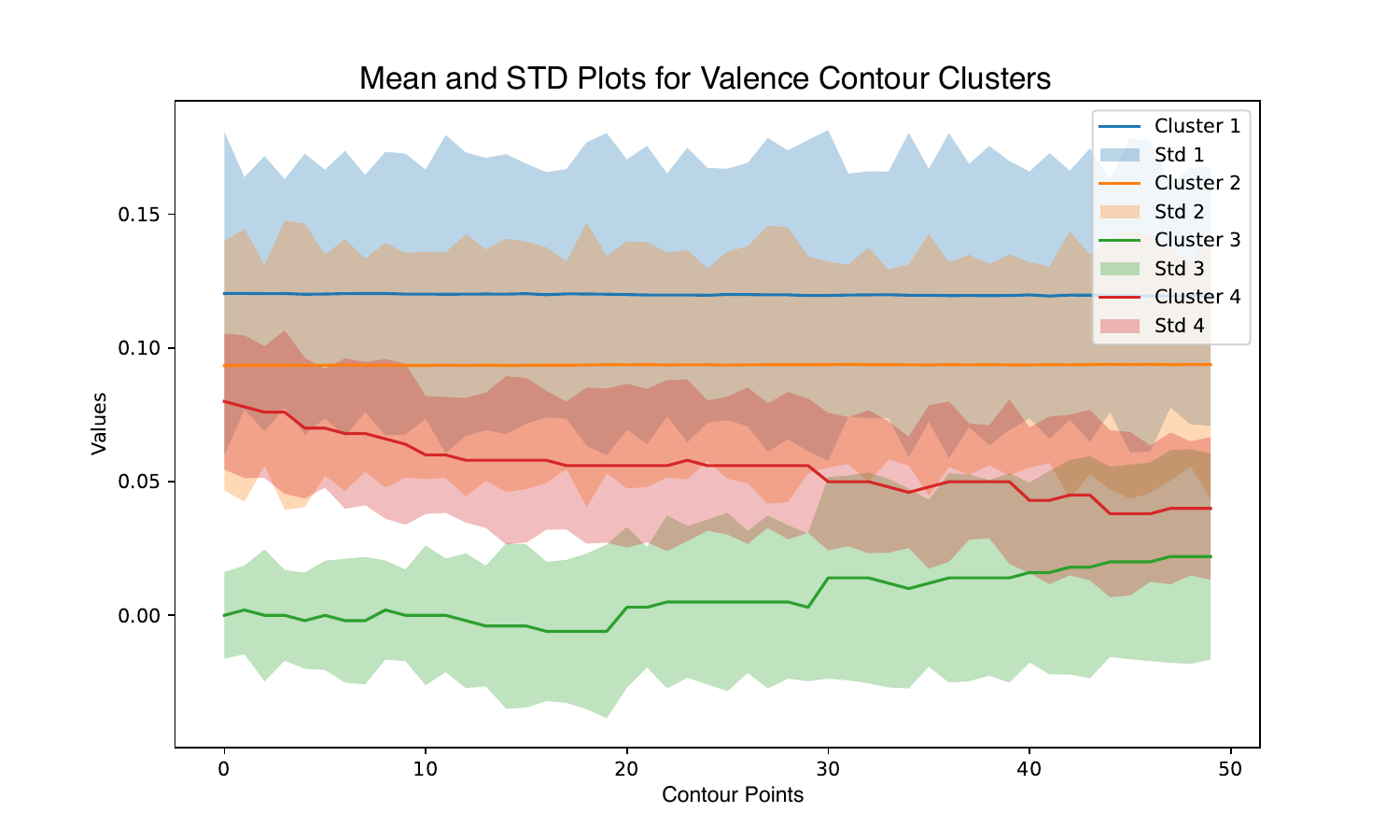}
    \caption{Valence}
    \label{fig:second}
  \end{subfigure}

  \caption{Arousal and valence contour mean and standard deviations for each C2P cluster}
  \label{fig:mean_std_contours}
\end{figure}

The self-supervised learning structure of the proposed C2P network discovers affect contour clusters that are highly predictable from speech. We investigate the $k=4$ cluster setting in our experimental evaluations. Figure~\ref{fig:mean_std_contours} presents the resulting affect contour clusters (mean and std contours) for arousal and valence attributes. For both attributes, flat (no-change), increase, and decrease trends can be observed from contour clusters. This sets an interesting observation of the discovered clusters.

\begin{figure}[t]
  \centering
  \includegraphics[width=0.4\linewidth]{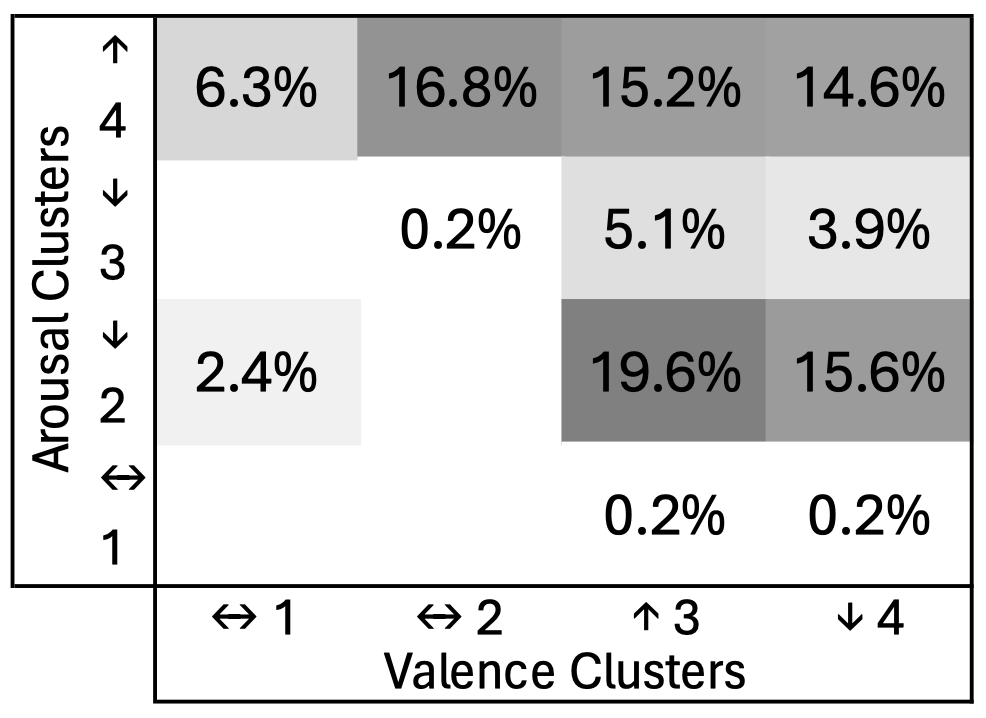}
  \caption{Affect cluster pairing percents for arousal and valence over the RECOLA dataset with contour cluster trends indicated as flat (no-change), increasing, or decreasing}
  \label{fig:pairings}
\end{figure}

Figure~\ref{fig:pairings} presents affect cluster pairing distributions (in percent scores) for arousal and valence over the RECOLA dataset with contour cluster trends indicated as flat, increasing, or decreasing. We can observe that the arousal contour~1 (A1) is a rear event, and valence contours~1~and~2 (V1, V2) have both flat (no-change) characteristics. On the other hand, the most dominant occurrences are i) A2-V3 with 19.6\% with a move towards south-east on the AV plane (calm), ii) A4-V2 with 16.3\% with a move towards north (increased arousal), iii) A2-V4 with 15.6\% with a move towards south-west (sad or bored), iv) A4-V3 with 15.2\% with a move towards north-east (happy), and v) A4-V4 with 14.6\% with a move towards north-west (angry).

\subsection{C2P Classification Results}

We investigate the performance of the C2P network with the classification performances of the $k$-class arousal and valence contour classification tasks. We also define two baseline classification systems compared to the proposed C2P network. The first baseline, affect-contour clusters (ACC), skips the AffectNet, and applies $k$-means clustering to the 50-unit long temporal affect attribute contours. The second baseline, average affect clusters (AAC), skips the AffectNet and applies $k$-means clustering to the mean affect attribute value of the 50-unit long temporal affect attribute contours. 

Classification performances for the proposed C2P, and baselines ACC and AAC models for arousal and valence attributes are given in Table~\ref{tab:av-classification}.

The proposed C2P model outperforms the baseline models in all metrics for arousal and valence attributes. When arousal and valence are compared to each other, we observe a better predictability for the arousal clusters. This aligns with the earlier results of speech-driven arousal and valence attribute prediction performances. Additionally, the results indicate that the baseline models show marginal improvements over chance levels. The AAC baseline model performs the worst among all three network structures. 

\begin{table}[th]
  \caption{Classification performances (accuracy, precision, recall, and F-score) for the proposed C2P, and baselines ACC and AAC models for arousal and valence attributes}
  \label{tab:av-classification}
  \centering
  \begin{tabular}{ l@{}l | c | c | c | c | c }
    \toprule
    \multicolumn{2}{c|}{\textbf{Model}} & 
    \multicolumn{1}{c|}{\textbf{Attrb}} & 
    \multicolumn{1}{c|}{\textbf{Acc}} & 
    \multicolumn{1}{c|}{\textbf{P}} &
    \multicolumn{1}{c|}{\textbf{R}} &
    \multicolumn{1}{c}{\textbf{F-score}} \\
    \midrule
    C2P & & A & 0.83 & 0.90 & 0.79 & 0.84 \\
    ACC & & A & 0.34 & 0.38 & 0.36 & 0.37 \\
    AAC & & A & 0.24 & 0.26 & 0.28 & 0.27 \\
    \midrule
    C2P & & V & 0.71 & 0.78 & 0.72 & 0.75 \\
    ACC & & V & 0.29 & 0.32 & 0.31 & 0.31 \\
    AAC & & V & 0.25 & 0.24 & 0.27 & 0.25 \\
    \bottomrule
  \end{tabular}
\end{table}

\section{Conclusions}
\label{sec:conclusions}

The literature has extensively studied predicting emotional attributes (regression) and emotional categories (classification) from speech. This remains an active area of research due to the inherent challenges in representing and quantifying emotions across different modalities, like speech. We propose a novel SSL-based cluster-to-predict (C2P) approach to address this. This approach aims to identify clusters of temporal affect contours that can be effectively predicted from speech representations. We evaluated our approach on the RECOLA dataset. The results demonstrate that affect-contour clusters learned using SSL with a target on predictability from speech achieve promising performance. In our four-class classification tasks, classification performances of the learned affect clusters achieved F1 scores of 0.84 for arousal and 0.75 for valence. Additionally, we observed dominant trends in the movement of arousal-valence contours across the four quadrants of the AV plane. This new perspective opens doors for future research on exploring novel emotion representations using SSL-based approaches.

%\bibliographystyle{IEEEbib}
%\bibliography{references}

\end{document}